\documentclass[preprint,proceedings]{rmaa}

\suppressfulladdresses 

\usepackage{paralist}

\usepackage{psfrag,color}

\def\spose#1{\hbox to 0pt{#1\hss}}
\def\gtsimm{\mathrel{\spose{\lower 3pt\hbox{$\sim$}}
        \raise 2.0pt\hbox{$>$}}}
  
\SetYear{2006}
\SetConfTitle{Massive Stars: Fundamental Parameters and Circumstellar Interactions}

\title{Radio observations of the massive stellar cluster Westerlund 1} 

\author{
  S. M. Dougherty,\altaffilmark{1}
  and J. S. Clark\altaffilmark{2}}

\altaffiltext{1}{National Research Council, Herzberg Institute for Astrophysics, Dominion Radio Astrophysical Observatory, Canada (sean.dougherty@nrc-cnrc.gc.ca).}
\altaffiltext{2}{Physics and Astronomy, Open University, Walton Hall, Milton Keynes, UK (jsc@star.ucl.ac.uk).}

\shortauthor{Dougherty \& Clark}
\shorttitle{RevMexAA(SC) Demo Document}

\listofauthors{S.M. Dougherty \& J.S. Clark}
\indexauthor{Clark, J.S.}
\indexauthor{Dougherty, S.M.}

\abstract{High-dynamic range radio observations of Westerlund~1 (Wd~1)
are presented that detect a total of 21 stars in the young massive
stellar cluster, the richest population of radio emitting
stars known for any young massive galactic cluster in the Galaxy. We
will discuss some of the more remarkable objects, including the highly
radio luminous supergiant B[e] star W9, with an estimated mass-loss
rate ($\sim 10^{-3}$~M$_\odot$~yr$^{-1}$) comparable to that of $\eta$
Carina, along with the somewhat unusual detection of thermal emission
from almost all the cool red supergiants and yellow hypergiants. There
is strong supporting evidence from X-ray observations that each of the
WR stars with radio emission are likely to be colliding-wind binaries.
}

\resumen{Se presentan observaciones de radio de alto rango din\'amico
hacia Westerlund 1 (Wd 1), que permiten detectar un total de 21
estrellas en el c\'umulo estelar joven y masivo, el cual tiene la
m\'as rica poblaci\'on de radioestrellas conocida dentro de los
c\'umulos estelares j\'ovenes gal\'acticos. Discutiremos algunos de
los objetos m\'as notables, incluyendo la estrella supergigante B[e]
identificada como W9, muy luminosa en radioondas, y que presenta una
tasa de p\'erdida de masa estimada ($\sim
10^{-3}$~M$_\odot$~yr$^{-1}$) comparable a la de $\eta$ Carina, junto
con la detecci\'on algo inusual de emisi\'on t\'ermica de casi todas
las supergigantes rojas e hipergigantes amarillas. Hay evidencia en
rayos X importante para proponer que cada una de las estrellas WR con
emisi\'on en radio es probablemente un sistema binario con vientos en
colisi\'on.
}
\addkeyword{Radio Continuum: Stars}
\addkeyword{Stars: Clusters: Westerlund 1}
\addkeyword{Stars: Individual: W9}

\begin{document}
\maketitle

\section{Introduction}
\label{sec:intro}
Wd~1 is a highly reddened ($A_v\sim12$~mag), compact
galactic cluster discovered by Westerlund \citep{Westerlund:1961}, and
now known to have a unique population of post-main sequence massive
stars with representative members of all evolutionary stages: OB
supergiants and hypergiants, red and yellow supergiants, and
Wolf-Rayet stars \citep{Clark:2002, Clark:2005}. With a total stellar
mass likely in excess of 10$^5$M$_{\odot}$ \citep{Clark:2005}, Wd~1 is
more massive than any of the other massive galactic clusters, and has
comparable mass to Super Star Clusters (SSC), previously identified
only in other galaxies.  If Wd~1 is an SSC within our own Galaxy, it
presents a unique opportunity to study the properties of a nearby SSC,
where it is possible to resolve the individual members of the cluster
population, and determine basic properties that are difficult in the
typically more distant examples.

\begin{figure}[!t]
  \includegraphics[width=\columnwidth, angle=0]{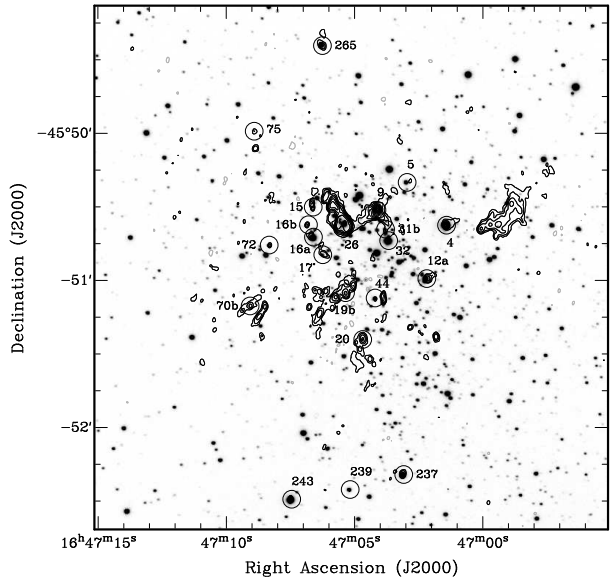}
\caption[]{8.6-GHz emission (contours) and a FORS R-band image
(greyscale). The radio sources with putative optical counterparts
listed are identified by the circles and corresponding Westerlund
numbers. \label{fig:radio_overlay}}
\end{figure}

Radio observations were obtained at four frequencies between 8.6 and
1.4~GHz with the Australia Telescope Compact Array. In addition, we
obtained an R-band image from the Very Large Telescope. The optical and
radio images were aligned by ensuring the peak of the optical and
radio emission of the brightest source at 8.6~GHz, W9, were
coincident. The resulting overlay is shown in
Fig~\ref{fig:radio_overlay}, and allows identification of the radio
emitting objects in Wd~1.

\section{Radio stars in Wd~1}
A total of 21 stars are associated with radio emission, making Wd~1
the richest population of radio emitting stars known for any young
massive galactic cluster, including the GC clusters \citep{Lang:2005}
and NGC~3603 \citep{Moffat:2002}. The stellar radio sources are blue,
yellow or red super- or hypergiants and WR stars, representative of
the different stages of massive star evolution.

The supergiant B[e] star W9 is by far the brightest stellar radio
emitter in the cluster, as anticipated from a previous radio
observation \citep{Clark:1998}.  Indeed, for a cluster distance of
4~kpc \citep{Kothes:2007}, W9 is one of the most radio luminous stars
in the Galaxy with $L_{\rm 8~GHz}=2\times10^{21}$ erg/s, similar to
the extreme LBV $\eta$ Car at radio minimum. The radio emission of W9
is resolved into a point source and an extended component, with the
former having a spectral index of $\sim+0.7$ and the extended
component having a flat spectrum consistent with optically-thin
thermal emission. This is interpreted as a stellar wind from the
underlying star surrounded by a more extended ejection nebula. The
estimated mass-loss rate is $\sim10^{-3}$~M$_\odot$/yr, unprecedented
for any massive star with perhaps the exception of an LBV during
outburst e.g. $\eta$ Car. The striking similarities of many of the
characteristics of W9 with $\eta$ Car, including the discovery of an
IR excess due to significant dust production, raises the intriguing
possibility that W9 is an LBV with an $\eta$ Car-like giant eruption
happening today.

Among the cooler RSG and YHG populations in Wd~1, 5/6 of the YHGs and
all four of the known RSGs are detected.  Each of these objects have a
spectral index consistent with optically-thin thermal emission. Being
too cool to ionize their own envelopes, the stellar wind material must
be ionized externally, most likely from the radiation field of the
cluster, but also potentially from hot companion objects, likely
another massive star. The mass of ionized material is typically
$\sim10^{-2}$~M$_\odot$ for the YHGs, but among the RSGs it is as high
as $\sim1$~M$_\odot$ for W26. Interestingly, the radio emission around
several of the RSGs appears to have an extended, cometary morphology
(e.g. W26) which may arise from ram pressure ablation due to a strong
cluster wind, similar to the process underlying the similar envelope
structure of the RSG IRS~7 in the Galactic Centre region
e.g. \citet{Zadeh:1991, Dyson:1994}.

Six of the 24 known WR stars \citep{Crowther:2006} have been detected
in this survey. For a mean radio derived mass-loss rate of
$4\times10^{-5}$~M$_\odot$~yr$^{-1}$ \citep{Leitherer:1997} a flux of
$(0.4-0.9)$~mJy is expected at a distance of 4~kpc. Even after taking
into account some degree of wind clumping, we should expect to detect
some WR stars in Wd~1. Each of the detected WR stars have flat
spectral indices, that we suggest results from a composite spectrum of
thermal and non-thermal emission, as often observed in colliding-wind
binary systems \citep[e.g][]{Dougherty:2000}. This hypothesis is
corroborated with Chandra observations of Wd~1 that show the WR stars
are typically X-ray bright with $L_x\sim10^{32-33}$ erg~s$^{-1}$ and
$kT>2.6$ keV i.e. $T>3\times10^7$~K \citep{Clark:2007}. Such
temperatures are expected in the post-shock flow of wind-wind
interaction shocks e.g \citet{Stevens:1992}.  The IR excess in the
WC-type star W239 (WR F) has been interpreted arising from dust in a
colliding-wind \citep{Crowther:2006} and together with evidence from
photometric and spectroscopic observations, it appears the WR binary
fraction in Wd~1 is unprecedently high, in excess of ~70\% \citep{Clark:2007}.


\begin{thebibliography} 

\bibitem[Clark et al. (2007)]{Clark:2007}
 Clark, J.S. et al. 2007, \aap, submitted

\bibitem[{{Clark} {et~al.}(2005)}]{Clark:2005}
{Clark}, J.~S., {Negueruela}, I., {Crowther}, P.~A., \& {Goodwin}, S.~P. 2005,
  \aap, 434, 949

\bibitem[{{Clark} \& {Negueruela}(2002)}]{Clark:2002}
{Clark}, J.~S. \& {Negueruela}, I. 2002, \aap, 396, L25

\bibitem[{{Clark} {et~al.}(1998)}]{Clark:1998}
{Clark}, J.~S., et al. 1998, \mnras, 299, L43

\bibitem[{{Crowther} {et~al.}(2006)}]{Crowther:2006}
{Crowther}, P.~A., {Hadfield}, L.~J., {Clark}, J.~S., {Negueruela}, I., \&
  {Vacca}, W. 2006, \mnras, 372, 1407

\bibitem[{{Dougherty} \& {Williams}(2000)}]{Dougherty:2000}
{Dougherty}, S.~M. \& {Williams}, P.~M. 2000, \mnras, 319, 1005

\bibitem[{{Dyson} \& {Harquist}(1994)}]{Dyson:1994}
Dyson, J.E., \& Hartquist, T.W. 1994, \mnras, 269, 447

\bibitem[{{Kothes} \& {Dougherty}(2007)}]{Kothes:2007}
{Kothes}, R. \& {Dougherty}, S.M. 2007, \aap, submitted

\bibitem[{{Lang} et al. (2005)}]{Lang:2005}
Lang, C.C., Johnson, K.E., Goss, W.M., \& Rodriguez, L.F. 2005, \aj, 130, 2185

\bibitem[{{Leitherer} {et~al.}(1997)}]{Leitherer:1997}
{Leitherer}, C., {Chapman}, J.~M., \& {Koribalski}, B. 1997, \apj, 481, 898

\bibitem[{{Moffat} {et~al.}(2002)}]{Moffat:2002}
{Moffat}, A.~F.~J. et al. 2002, \apj, 573, 191

\bibitem[Stevens et al. (1992)]{Stevens:1992}
Stevens, I.R., Blondin, J.M., Pollock, A.M.T. 1992, \apj, 286, 265

\bibitem[{{Westerlund}(1961)}]{Westerlund:1961}
{Westerlund}, B. 1961, \pasp, 73, 51


\bibitem[Yusef-Zadeh \& Morris (1991)]{Zadeh:1991}
Yusef-Zadeh, F., \& Morris, M. 1991, \apj, 371, 59

\end{thebibliography}
\end{document}